# Nuclear magnetic resonance measurements reveal the origin of the Debye process in monohydroxy alcohols

C. Gainaru,[1,*] R. Meier,[2] S. Schildmann,[1] C. Lederle,[1] W. Hiller,[3] E. A. Rössler,[2] R. Böhmer[1,#]

[1] *Fakultät für Physik, Technische Universität Dortmund, 44221 Dortmund, Germany*
[2] *Physikalisches Institut, Universität Bayreuth, 95440 Bayreuth, Germany*
[3] *Fakultät für Chemie, Technische Universität Dortmund, 44221 Dortmund, Germany*

Monohydroxy alcohols show a structural relaxation and at longer time scales a Debye-type dielectric peak. From spin-lattice relaxation experiments using different nuclear probes an intermediate, slower-than-structural dynamics is identified for n-butanol. Based on these findings and on diffusion measurements, a model of self-restructuring, transient chains is proposed. The model is demonstrated to explain consistently the so far puzzling observations made for this class of hydrogen-bonded glass forming liquids.

The dielectric response of monohydroxy alcohols, an important class of hydrogen-bonded liquids, was treated already by Peter Debye using a model which he reviews in his 1929 book [1]. Later, it was recognized that these alcohols, in addition to a very intense Debye process, exhibit two relaxations at higher frequencies [2,3]. The latter two are, by now, undisputedly identified as the structural and the Johari-Goldstein relaxation [4]. However, the origin of the pronounced Debye peak at low frequencies, often linked to the presence of hydrogen bonds and discussed in terms of supramolecular structures, has remained subject to stimulating controversy. These structures manifest themselves as prepeaks in the static structure factor [5], likely due to hydrogen bonded chains [6,7,8] that are most directly identified from molecular dynamics simulations [9,10]. It is remarkable that despite continued efforts the Debye process has resisted observation using, e.g., calorimetric [11] and viscoelastic experiments [12]. The idea to increase the amplitude of the Debye peak by increasing the number of hydrogen bonds fails badly: The peak vanishes altogether if two or more OH groups are in proximity on the same alcohol molecule. It adds to the complexity of the situation that $H_2O$, often discussed in analogy to alcohols, does also show a Debye-like process despite the presence of effectively two hydroxyl groups per molecule [13]. Not only the similarity to the dielectric absorption of water, but also the importance of "simple" alcohols as solvents calls for an understanding of their microscopic dynamics.

On the basis of spin-relaxation times $T_1$ for n-butanol, a well studied glass former [3,6,8,9,14], we provide evidence for a slower-than-structural hydroxyl group dynamics using nuclear magnetic resonance (NMR). Taking into account also self-diffusion experiments and dielectric measurements a transient chain model is developed which rationalizes all the perplexing features of alcohols just summarized, thus resolving a long standing puzzle.

Fig. 1 presents $^1$H spin-lattice relaxation rates of n-butanol which was specifically isotope labeled either at its OH site [$CD_3$-$(CD_2)_3$-OH] or at its alkyl part [$CH_3$-$(CH_2)_3$-OD] [15]. From data recorded at a Larmor frequency of $\omega_L/2\pi$ = 55.6 MHz the rate maxima for the two species are seen to appear at temperatures about $\Delta T \approx 28$ K apart. Similar observations can be made from $^2$H-$T_1$ measurements, except that here due to the larger NMR coupling constant the $^2$H rates are shifted to larger values. Both probes indicate that, at a given T, the underlying molecular correlation times $\tau_{OH}$ are much slower than $\tau_{alkyl}$ [16,17].

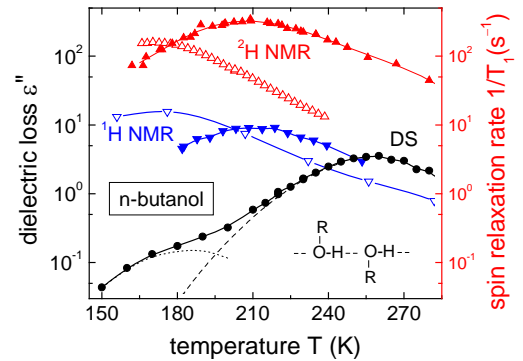

Fig. 1 (Color online) Spectral density of n-butanol as probed using the dielectric loss (●, 50 MHz) and NMR: $^1$H-$T_1$ (▼, $\omega_L/2\pi$ = 55.6 MHz; ▽, 46.0 MHz) and $^2$H-$T_1$ (▲, △, 46.2 MHz). The closed triangles refer to measurements at the hydroxyl group, the open triangles to those at the alkyl part of butanol. The solid lines are drawn to guide the eye. The dashed line marks the contribution of the Debye-process, and after its subtraction from the total loss the dashed curve, corresponding to the α-process, is obtained. The inset shows a sketch of two hydrogen bonded alkanols where R designates an alkyl group.

Fig. 1 includes dielectric losses $\varepsilon''(\omega_L)$ measured in the T range of the Debye peak from which the time constant $\tau_D = 1/(2\pi\nu_{peak})$ can be read off. By subtracting the contribution of the Debye relaxation, similar to previous work [14], a peak due to the structural relaxation is obtained from which $\tau_\alpha$ can be gained. From Fig. 1 it is clear that $\tau_D > \tau_{OH} > \tau_\alpha$, thus an additional intermediate time scale exists, providing clues to identify the microscopic nature of the Debye process.

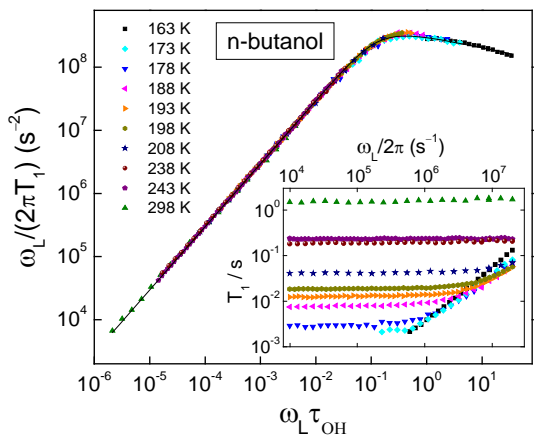

Fig. 2 (Color online) The inset shows proton spin-lattice relaxation times ($^1$H-$T_1$) of supercooled $CD_3$-$(CD_2)_3$-OH as measured in a broad range of T and $\omega_L$. The main frame presents the same data in a scaled susceptibility format. The only variable required to obtain this virtually perfect masterplot is the T dependent correlation time $\tau_{OH}$. The solid line is a fit based on a Cole-Davidson spectral density

For a detailed comparison with $\varepsilon''$, it is essential to determine also $T_1$ in broad $\omega_L$- and T-ranges. Our present field cycling NMR experiments [18] on [$CD_3$-$(CD_2)_3$-OH] allowed us to vary $\omega_L$ by more than 3 decades, see the inset of Fig. 2. These data were replotted in the susceptibility format, $\chi''(\omega) \propto \omega/T_1$. From the $\chi''$ peaks at the lowest T the correlation times $\tau_{OH}$ were determined on the basis of a Cole-Davidson spectral density. We obtained a very good fit employing an exponent $\beta_{CD} = 0.22$ which implies a very broad distribution of time scales [18]. The vertical shift required to obtain the practically perfect master curve shown in Fig. 2 yielded $\tau_{OH}$ for higher T. These $\tau_{OH}$ and a data point measured at room temperature [19] are collected in an Arrhenius diagram, Fig. 3, together with the time constants $\tau_D$ and $\tau_\alpha$ from dielectric spectroscopy (DS). The OH correlation times $\tau_{OH}$ are ~ 8 times longer than $\tau_\alpha$, and ~ 50 times shorter than $\tau_D$ for T ~ 160 K.

With one proton per molecule, $^1$H-$T_1$ measurements are solely sensitive to *inter*molecular dynamics while $^2$H is sensitive to the reorientation of a single hydroxyl group. Therefore, at first glance, it appear surprising that $^1$H-$T_1$ *and* $^2$H-$T_1$ yield practically the same $\tau_{OH}$. However, this similarity is to be expected if the polar head groups aggregate in a manner sketched in Fig. 1: Then, a typical OH bond direction is oriented approximately along the vector connecting two adjacent hydroxyl protons [20]. Thus, our measurements support the time honored notion that alcohols form chain-like structures, see, e.g., Ref. 8. The motion of the nonpolar alkyl groups about the "chain" backbone can be mapped out via $^{13}$C-NMR. From $T_1$ measurements at different carbon sites we determined $\tau_{CH2} = \tau_{alkyl}$ and find $\tau_{alkyl} \approx \tau_\alpha$ [15] in accord with results for propanol [4,16]. This demonstrates that, with the nomenclature known from polymers, the $\alpha$-process is to be viewed as a segmental motion.

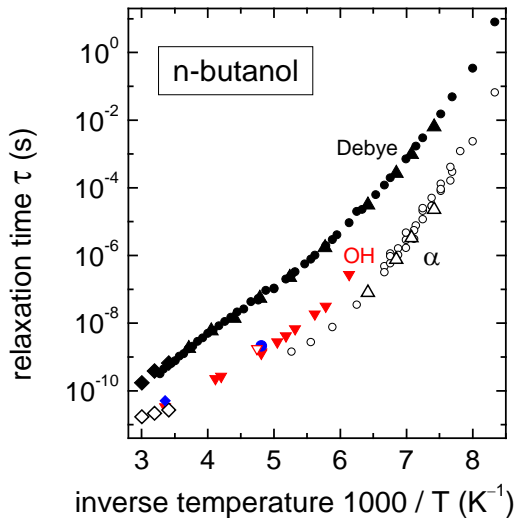

Fig. 3 (Color online) Arrhenius plot of n-butanol including time constants from DS (some low-frequency data are from Ref. 14) and from NMR. The present data ($\tau_D$, ●; $\tau_\alpha$, ○; $\tau_{OH}$ from $^1$H-$T_1$,▼,▽; $\tau_{OH}$ from $^2$H-$T_1$,●) is compared with literature values for $\tau_D$ (▲, Ref. 8; ◆, Ref. 3), $\tau_\alpha$ (△, Ref. 2; ◇, Ref. 3) and $\tau_{OH}$ (◆; Ref. 19)

From measurements of the self-diffusion coefficient D of n-butanol [15] and from those of its viscosity η [21], for 318 K > T > 288 K we determined the hydrodynamic radius of the translationally moving moieties to be $r_H = k_B T/(6\pi\eta D) \approx$ 2.3 Å. This is in good agreement with the molecular radius estimated for butanol [22] ruling out a supramolecular transport. A chain motion without displacing the chain as a whole is pictured in Fig. 4. For simplicity only a single, relatively short chain is highlighted. The sequence of frames in this figure visualizes a snake-like motion induced by a successive loss (or gain) of segments at its one end and a gain (or loss) of segments at its other end. The "core" of the chain, i.e., the "inner" hydrogen-bridged hydroxyl groups are temporarily held together by electrostatic forces which in turn are responsible for the segregation of the polar from the nonpolar groups.

We now show that in such a transient-chain scenario a slow dipolar, single-exponential, i.e., Debye-type process naturally arises. The succession of the polar OH groups along the curvilinear chains yields a supramolecular dipole moment, similar to what happens in type A polymers [23]. The reorientation of this moment, i.e., that of the chain's end-to-end vector, which we thus associate with $\tau_D$, is much slower than $\tau_{OH}$. Since the de- and attachment of segments is a stochastic process, chain length variations arise as time progresses. However, with $\tau_D \gg \tau_{OH},\tau_\alpha$ only the mean length matters on the time it takes to reorient the end-to-end vector. This precludes a time scale polydispersity due to chain length variations. Furthermore, during $\tau_D$ the chain effectively averages over the fluctuations in its environment. Taken together these effects take place on a time scale of the order of $\tau_\alpha$ leading to a single exponential Debye relaxation.

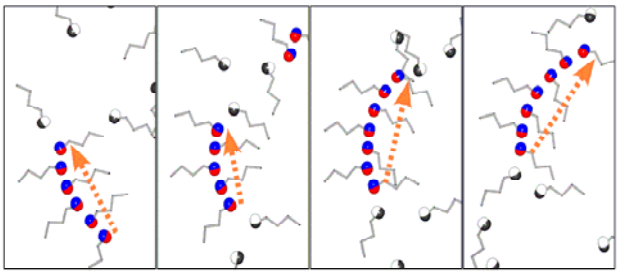

Fig. 4 (Color online) Schematic illustration of the transient chain model. Mutually bonded OH groups are shown in color. The sequence of frames is meant to visualize how molecules attach to the chain and detach from it. The dotted arrows highlight the end-to-end vector of the self-restructuring, transient chain. Its reorientation, corresponding to the Debye process, is obviously very slow on the scale set by the elementary steps. The chain-length fluctuations are much faster than $\tau_D$ leading to an exponential relaxation. An animated version of this figure is available from the authors upon request.

One may designate $\tau_D$ as $\tau_\parallel$ (referring to parallel to the chain) and $\tau_\alpha$ as $\tau_\perp$. Taking the polymer analogy one step further, the molecular dipole moment $\mu$ can be thought to be decomposed into components locally oriented along the chain ($\mu_\parallel$) and perpendicular to it ($\mu_\perp$). Of course, $\mu$ is larger than either projection. The $\mu_\parallel$ components add up leading to a Debye process with enhanced intensity. For the "segmental" motion, which we associate with the α-process, $\mu_\perp < \mu$ yields a reduced relaxation strength as experimentally observed [24]. Furthermore, from our model $\Delta\varepsilon_D/\Delta\varepsilon_\alpha$ and $\tau_D/\tau_\alpha$ both are expected (i) to decrease for shorter chains and (ii), also in agreement with experimental observations, to be largely independent of the length of the alkyl rests. The latter essentially act as spacers between the transient, permanently self-reconstructing chains.

To visualize the situation, comparisons with type A polymers can be helpful but they should not be overestimated because in our case higher-order end-to-end (Rouse-type) normal modes based on the existence of *permanent* links between chain segments are absent [23]. It may be asked how long the transient alcohol chains typically are. By assuming that "free" molecules are transported to and away from the chain ends on a time scale not too different from $\tau_\alpha$ the life time of a molecule within a chain with N segments is of the order $N\tau_\alpha$. On the other hand, one may consider the orientation of an OH group as essentially fixed as long as it is part of a chain, implying that $\tau_{OH}$ is roughly of the order of $N\tau_\alpha$.

At first glance, the sketch in Fig. 4 seems related to recent proposals [10,25]. But the Debye peak and the structural relaxation are assigned in entirely different ways by the previous approaches. Nevertheless, being based on a similar intermolecular association, they are expected to capture geometrical aspects analogously, e.g., when rationalizing trends in the dispersion strengths at high pressures or when the OH group is screened either by moving it away from a chain end or by considering branched hydrocarbon chains. This implies the growth of $\Delta\varepsilon_\alpha$ at the expense of $\Delta\varepsilon_D$ not only when pressure [25,26], but also temperature is increased. Our spectral analysis (not shown) confirms this nontrivial $\Delta\varepsilon_\alpha(T)$ behavior.

The present model also explains why relaxation modes corresponding to a Debye peak are not observed in mechanical spectroscopy. Due to the transient nature of the chains, the motion of the end-to-end vector does not couple to changes of the internal stress field. Near $\tau_D$ a mechanical relaxation mode is thus not expected. All but a minute decrease of the overall stiffness could result when reducing the chain length, e.g., by pressure application. Furthermore, the "wandering snakes" do lead neither to energy fluctuations nor to entropy fluctuations on the time scale set by $\tau_D$. However, a minor calorimetric signature corresponding to time scales $\tau_{OH} > \tau_\alpha$ could show up [27]. Only very few methods apart from DS are sensitive to transient, dipolar chains. The only other *direct* evidence for dynamics on the $\tau_D$ scale, that we are aware of, is from largely overlooked Kerr-effect work on monoalcohols [28].

For polyalcohols, with each molecule carrying several OH groups, branched structures result. Hence, a well-defined end-to-end vector does not exist. Merely, a component dipole moment may be associated with each temporarily existing branch. However, the unsynchronized OH switching occurring in such a multiple-branch structure will lead to an effective mutual cancellation of the component moments. Thus, with a vanishing net dipole moment a supramolecular dielectric signature does not arise from such branched structures and a Debye process is not expected in harmony with the experimental situation.

Since the transient chain model proposed herein on the basis of dielectric and NMR data coherently captures all essential features of the so far puzzling observations made for the dynamics of monoalcohols, it is tempting to check which modifications of the model are required to render it applicable also to water with its Debye-like dielectric relaxation. A one-to-one correspondence to alcohols is not obvious, since $H_2O$'s local bonding is predominantly fourfold in character. Essentially chain-like features were nevertheless identified from molecular-dynamics simulations of water [29]. It remains to be seen how the cohesive forces arise which seem necessary to support the persistence of such chain-like structures in water.

We thank A. Geiger and H. Weingärtner for stimulating discussions and A. Nowaczyk for his help with Fig. 4. Partial support of this project by the Deutsche Forschungsgemeinschaft under Grant No. BO1301/8-1 is gratefully acknowledged.

———————————

Corresponding authors:
*catalin.gainaru@uni-dortmund.de
#roland.boehmer@tu-dortmund.de